# To Lane or Not to Lane? Comparing On-Road Experiences in Developing and Developed Countries using a New Simulator "RoadBird"

**Md. Masum Mushfiq**
Lecturer, Department of Computer Science and Engineering
Bangladesh University of Engineering and Technology, Dhaka-1205, Bangladesh
Email: masummushfiq@cse.buet.ac.bd

**Tarik Reza Toha**
Lecturer (Computer), Department of Textile Machinery Design and Maintenance
Bangladesh University of Textiles, Dhaka-1208, Bangladesh
Email: toha@butex.edu.bd

**Saiful Islam Salim**
Graduate Student, Department of Computer Science and Engineering
Bangladesh University of Engineering and Technology, Dhaka-1205, Bangladesh
Email: 1018052067@grad.cse.buet.ac.bd

**Aaiyeesha Mostak**
Graduate Student, Department of Computer Science and Engineering
Bangladesh University of Engineering and Technology, Dhaka-1205, Bangladesh
Email: 0419052029@grad.cse.buet.ac.bd

**Masfiqur Rahaman**
Undergraduate Student, Department of Computer Science and Engineering
Bangladesh University of Engineering and Technology, Dhaka-1205, Bangladesh
Email: 1505111.mr@ugrad.cse.buet.ac.bd

**Najla Abdulrahman Al-Nabhan**
Assistant Professor, Computer Science Department
King Saud University, Riyadh, Saudi Arabia
Email: nalnabhan@ksu.edu.sa

**Arif Mohamin Sadri**
Assistant Professor, Moss Department of Construction Management
Moss School of Construction, Infrastructure & Sustainability
Florida International University, Miami, Florida-33174, USA
Email: asadri@fiu.edu

**A. B. M. Alim Al Islam**
Professor, Department of Computer Science and Engineering
Bangladesh University of Engineering and Technology, Dhaka-1205, Bangladesh
Email: alim_razi@cse.buet.ac.bd

Word Count: 6605 words + 3 table = 7355 words

## ABSTRACT

Even though the traffic systems in developed countries have been analyzed with rigor and operated efficiently, the same does not generally hold for developing countries due to inadequate planning, design, and operations of their transportation systems. Because of inherent differences between internal infrastructures, the systems deployed in developed countries may not be amenable to developing ones. Besides, the traffic systems of developing countries are not well-studied in the literature to the best of our knowledge. For example, it is yet to explore how a developed country's lane-based traffic flow would perform in the context of a developing country, which generally experiences non-lane-based traffic. As such, by using our newly developed traffic simulator 'RoadBird', we investigate outcomes of both lane-based and non-lane-based traffic from the contexts of both developing and developed countries. To do so, we run simulations over real road topologies (extracted from the GIS maps of major cities such as Dhaka, Miami, and Riyadh) considering different scenarios such as lane-based or non-lane-based flows, homogeneous or heterogeneous traffic, with or without pedestrians, etc. We also incorporate different car-following and lane-changing models to mimic traffic behaviors and investigate their performances. While the lane changing dilemma remains an open research question, our experimental evidences indicate: (i) lane-based approaches will not necessarily perform better in the case of currently-adopted non-lane-based scenarios; and (ii) non-lane-based strategies may benefit system performance in lane-based scenarios while having heavy mixed traffic. Nonetheless, we reveal several new insights for on-road experiences both in developing and developed countries.

## INTRODUCTION AND MOTIVATION

Dhaka, the capital of Bangladesh, is one of the mega-cities of the world. However, the chaotic traffic system of Dhaka is a major headache for the city dwellers. Traffic congestion in Dhaka is responsible for wasting around 3.2 million working hours daily costing the economy billions of dollars [1]. In this condition, saving even a single minute on an average can save millions of dollars. Traffic simulation software is a useful tool for system-wide traffic impact assessment and sustainable policymaking that can come into play in cases such as Dhaka. This can be done by testing the impact of a proposed policy on the intended traffic network. However, the effectiveness and applicability greatly vary with how accurately the simulator can mimic the desired traffic stream. As Dhaka is a city of unstructured (non-lane based) heterogeneous (high mix of slow and faster-moving vehicles) traffic, its traffic stream experiences diversified on-road scenarios such as pedestrian on road, illegal parking, deliberate rule violation, jaywalking, and so forth, which are quite different from the structured (lane-based) traffic systems of developed countries with homogeneous traffic in terms of speed. Existing traffic simulators lack the ability to accommodate such on-road scenarios, and hence, the need for an advanced simulator with more customized features comes into play for simulating both structured and unstructured traffic behaviors.

As already pointed, vehicles not following lanes are one of the major attributes of Dhaka city's traffic. On the contrary, the traffic stream is mostly lane-based in the major cities of developed countries. Therefore, it is quite natural for the following question to arise: "What would happen if the traffic system of Dhaka is converted into a lane-based system?" Likewise, what would happen if the major cities in developed countries adapt to non-lane-based strategies? In this study, we try to investigate this dilemma between "to lane or not to lane" to decide on better system performance. To the best of our knowledge, such a study is yet to be done in the literature, and we are the first to do so.

In road to performing our study, we have developed a new microscopic traffic simulator named RoadBird, which is based on its earlier version formerly named DhakaSim [2]. In our current work, we extend the simulator to simulate both lane-based and non-lane-based traffic along with incorporating different traffic behavioral phenomena such as car-following and lane-changing models to mimic the realistic traffic behavior. We investigate their performance and choose the best applicable model to conduct further experiments. The current version of RoadBird can also simulate behaviors of other traffic entities such as pedestrians, slow vehicles (for example rickshaws), bikes, etc. In this study, we run simulations on topologies extracted from the GIS map of Dhaka, Miami, and Riyadh to analyze their performances and trajectories. After that, we use different performance metrics to measure performances in different scenarios.

Based on our work, we make the following set of contributions:

- We present a new traffic simulator RoadBird, which is capable of simulating both lane-based and non-lane-based traffic in the presence of both homogeneous and heterogeneous traffic streams. RoadBird can also take account of diversified scenarios experienced over roads in developing countries. Examples include pedestrian on road, movement of slow vehicles such as rickshaws, etc.
- We simulate RoadBird to compare the performances in both the contexts of developed and developing countries by varying traffic load on the network from low to high vehicle generation rate. We also compare the performances by varying the ratio of different types of vehicles i.e., vehicular mix. Here, we use the road networks extracted from Miami and Riyadh as representatives of developed country's road network, and road network extracted from Dhaka as a representative of developing country's road network
- According to our observations, we further analyze and show that non-lane-based systems exhibit better performance by utilizing the road space more efficiently when the traffic is heterogeneous and roads are narrow. However, when the traffic is homogeneous and roads are wide, lane-based systems exhibit better speed and flow rate of vehicles.

- While lane-based approaches will not necessarily yield in more efficient outcomes in non-lane-based scenarios, the other side of the coin is not this impotent. This happens as non-lane-based strategies may benefit system performance in lane-based scenarios with heavy mixed traffic, which we confirm by our analysis.

## BACKGROUND AND RELATED WORK

Dhaka is one of the mega-cities of the world. However, life in Dhaka is difficult due to the chaotic traffic condition of the city. Microscopic traffic simulation is widely used and one of the most effective ways to predict traffic behavior. These tools can aid in taking important transportation engineering decisions by simulating the proposed decisions and analyzing its impact on the existing traffic network. However, its effectiveness greatly depends on the accuracy of mimicking the intended traffic pattern. Although there exist several microscopic traffic simulators in the literature, they fail to mimic the non-lane-based heterogeneous, i.e., unstructured traffic stream of Dhaka city. Hence, we need a customized traffic simulator to simulate the impact of a transportation engineering policy on an unstructured traffic stream. Using this custom simulator, we can investigate the impact of converting a non-lane-based road network with heterogeneous traffic into a lane-based road network.

Existing state-of-the-art traffic simulators such as VISSIM [3], MITSIM [4], and SUMO [5] mainly focus on lane-based homogeneous, i.e., structured traffic stream. However, in developing countries, the traffic stream is often non-lane-based and heterogeneous, i.e., unstructured. Hence, it is hardly possible to simulate the traffic behavior of developing countries with existing traffic simulators. In our previous work, we have developed a non-lane-based traffic simulator named as DhakaSim [2] to simulate the diversified behavior of traffic. However, it lacks the capability of simulating a lane-based structured traffic stream. Moreover, no comparative study between lane-based and non-lane-based heterogeneous traffic performance has been done. Hence, we extend the existing DhakaSim traffic simulator to RoadBird to simulate both lane-based and non-lane-based traffic streams and conduct an extensive experimental study to derive a conclusive remark about "To lane or not to lane". Next, we present some existing work on traffic simulations and related models.

Fellendorf et al., [3] propose a microscopic behavior-based multi-purpose traffic simulator VISSIM. It offers a wide variety of urban and highway applications through integrating public and private transportation. It is the ideal tool for state-of-the-art transportation planning and operations analysis. The latest version of VISSIM supports heterogeneous traffic that follows lane-based discipline. It also includes pedestrian movement and car-parking, however, in a structured manner. Huang et al., [6] discuss how VISSIM can be used to simulate for safety criterion. Simulated conflicts generated by the VISSIM simulation model and identified by SSAM [7] are compared to the traffic conflicts measured in the field. They also test the prediction performance of the conflict prediction model. Aghabayk et al., [8] give an overview of car-following and lane-changing models of VISSIM and show how to calibrate VISSIM using multithreading. Ben-Akiva et al., [4] propose another advanced traffic simulation tool named MITSIM that evaluates the impacts of alternative traffic management system designs, traveler information systems, public transport operations, and various ITS strategies at the operational level and assists in their subsequent refinement. However, MITSIM supports lane-based homogeneous traffic only.

Vedagiri et al., [9] propose a simulator named HETEROSIM to simulate heterogeneous traffic flow considering Indian road traffic. They estimate the saturation flow rate of heterogeneous traffic considering the effect of road. They use passenger car units (PCU) per unit width of the road as a unit of saturation flow. From the simulation result, they have found a linear relationship between saturation flow (PCU/m) and road width (m). Arsasan et al., [10] measure one of the fundamental characteristics of traffic flow, i.e., concentration using HETEROSIM. They argue that the traditional concept of concentration cannot be applied to heterogeneous and non-lane-based traffic and propose a new concept named as area-occupancy to measure traffic concentration for heterogeneous and non-lane-based traffic.

There are few studies about modeling heterogeneous traffic flow. Arasan et al., [11] propose a simulation framework for the traffic-flow model that is prepared in such a way that the absence of lane discipline in mixed traffic flow conditions is taken into account. Common issues related to traffic simulation such as vehicle generation, logics for vehicular movement, etc., are also described in detail in the context of heterogeneous traffic conditions. A different car-following model is discussed by Olstam et al., [12], however, all are for lane-based traffic. Jin et al., [13] propose a non-lane based full velocity difference car-following model where they incorporate the lane width effect in car-following models and show that lateral separation effect greatly enhances the realism of non-lane based car-following models. Muniruzzaman et al., [14] propose a method to calibrate and validate non-lane based microscopic simulation models, however, they do not compare between lane-based and non-lane based traffic systems.

Ngoduy et al., [15] investigate how different the car-following type influences the overall stability of heterogeneous traffic flow dynamics using a generalized multi-class car-following model. Yang et al., [16] propose a cellular automata-based traffic flow model for single lane traffic. The proposed model discriminates the four types of car–truck combinations. These works only focus on motorized vehicles such as cars, and trucks. However, in a developing country like Bangladesh, India, Kenia, etc., the road having non-motorized vehicles are very common. Besides in these countries, human mobility is very common on the road in a random pattern. Therefore, pedestrians play a vital role in traffic flow dynamics in a heterogeneous traffic condition. Nokandeh et al., [17] demonstrate the dynamic nature of PCU (Passenger Car Unit) factors on two-lane intercity highways under highly heterogeneous traffic composition. The main limitation of their model is that they adopt lane-based discipline, which is very rare in a developing country. Li et al., [18] propose a heterogeneous car-following model of low and high sensitivity vehicles using the linear stability theory. They only consider two types of vehicles not pointing any heterogeneous non-motorized traffic.

**Mathew** et al., [19] propose a space discretization–based simulation framework named as SiMTraM to address the driver behavioral models in the heterogeneous traffic stream. Here, the lane is divided into multiple strips. With this model, continuous lateral movement can be modeled by defining very small strip widths. This simulator can simulate both the lane-based and non-lane-based traffic. However, the proposed simulator has no implementation for the random movement of the pedestrians and non-motorized traffic on the road. Agarwal et al., [20] propose an agent-based framework, that uses a queuing model to simulate the mobility of only motorized vehicles such as cars, bikes, and motorbikes. Mohan et al., [21] propose a parsimonious model of heterogeneous traffic that can capture the unique phenomena of the gap-filling behavior of vehicles. To capture the effect of heterogeneous and non-lane-based traffic, they extend an existing second-order continuum model of traffic flow using area occupancy for traffic concentration instead of density. They have calibrated and validated the model using field data from an arterial road in Chennai city. Chand et al., [22] develop a dynamic PCU for the candidate signalized intersections catering to mixed traffic conditions in Indian cities. However, the proposed simulator does not deal with random movement of the pedestrians and non-motorized traffic on the road.

All of the aforementioned studies explore different aspects of lane-based structured traffic. However, a comparative study between the lane and non-lane-based traffic with heterogeneous motorized and non-motorized traffic stream is yet to be done. Hence, we perform comparative study among structured and unstructured traffic through our custom designed and developed simulation software named as RoadBird in this work. Next, we present the methodology of our study.

## METHODOLOGY

In our study, we compare the performance of lane-based and non-lane-based road networks. We present the flowchart of our comparative study in **Figure 1.**

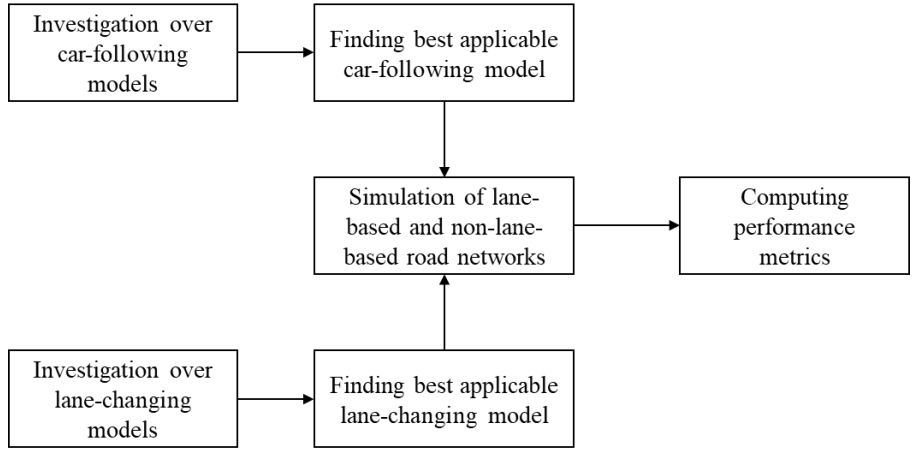

**Figure 1: Flowchart of our comparative study**

### A. Car-following Models

In our study, we implement three different car-following models and compare the performance by varying these models. Next, we describe each of them.

1. *Newtonian car-following model with hard braking:* The first model is a simple model. In this model, the following equations are used to compute the speed, distance, and acceleration of a vehicle.

$$\Delta x = x_{n-1}(t) - s_{n-1} - x_n(t) \tag{1}$$

$$v_n(t+\tau) = \begin{cases} v_n + a_n\tau, & accelerating \\ \dfrac{\Delta x}{\tau}, & braking \end{cases} \tag{2}$$

$$x_n(t+\tau) = x_n(t) + v(t)\tau \tag{3}$$

$$a_n(t+\tau) = a \tag{4}$$

where, $\Delta x$ is the safe distance between leader and subject vehicle, $\tau$ is the time step, which is 1s for our simulation, $s_{n-1}$ is the effective length of vehicle $n-1$, $x_{n-1}(t)$ and $x_n(t)$ are the distance of the leading vehicle $n-1$ and subject vehicle $n$ in a link.

2. *Gipps' car-following model:* In the second model, distance, and acceleration are calculated as the naive model. However, speed is calculated according to Gipps' model [3] as shown in **Equation 5**.

$$v_n(t+\tau) = min \begin{cases} v_n(t) + 2.5a_n\tau\left(1 - \dfrac{v_n(t)}{v_n^d}\right)\sqrt{0.025 + \dfrac{v_n(t)}{v_n^d}} \\ b_n\tau + \sqrt{b_n^2\tau^2 - b_n[2(\Delta x) - v_n(t)\tau - \dfrac{v_{n-1}(t)^2}{b}]} \end{cases} \tag{5}$$

where, $\tau$ is the reaction time; $v_n(t)$ and $v_{n-1}(t)$ are the speed of the following vehicle $n$ and the leading vehicle $n-1$ at time step $t$, respectively; $v_n^d$ is the vehicle n desired speed; $a_n$ is the vehicle $n$ maximum acceleration; $b_n$ and $b$ are the most severe actual time gap which can be computed as, braking that the driver of vehicle $n$ wishes to undertake and the expected leading vehicle maximum deceleration, respectively. Here, no explicit braking is applied. So, vehicles collide with each other.

3. *Hybrid model:* Hybrid model is the combination of the previous two models. In this model, vehicles normally move with speed from Gipps' model **Equation 5**, however, if they are about to collide, braking is applied using **Equation 3**.

## B. Discretionary Lane Changing Models

Discretionary lane changing (DLC) models have 3 parts. They are:

- Desire to change the current lane
- Ensuring lane change is feasible
- Decision to change lane based on gap acceptance

We implement 3 DLC models in our simulation. Next, we describe each of them.

1. *Straight-forward Model:* In the naive model, when a vehicle cannot move forward or has a slower leader in proximity, it wants to change its lane. If its target lane has enough space to accommodate it, it just shifts to the target lane.

2. *Gipps' Model:* In the Gipps' model, we first compute the braking of the subject vehicle using **Equation 6**.

$$b_n(t) \quad = \quad v_n(t-1) - v_n(t) \tag{6}$$

$v_n(t-1)$ and $v_n(t)$ are computed using Gipps' formula from **Equation 5** where the subject is vehicle $n$ and target leader is vehicle $n-1$. Similarly, the braking of the target follower is computed using **Equation 6**, where the target follower is vehicle $n$ and subject is vehicle $n-1$. Now, if computed braking of the subject and the target follower is lower than their maximum desired braking, lane change is feasible. Gap acceptance probability is calculated using **Equation 7**.

$$p(t) \quad = \quad \begin{cases} 1 - e^{-\lambda(t-T)}, & t > T \\ 0, & otherwise \end{cases} \tag{7}$$

where, $\lambda$ is a co-efficient, $T$ is the critical time gap, $t$ is the actual time gap which can be computed as, $t = \frac{g}{v_n}$. Here, $g$ is the lead/lag gap, $v_n$ is the speed. The probability that the gap is accepted is the product of the probability that the lead gap is accepted and the probability that the lag gap is accepted, that is,

$$p(t^{lead}, t^{lag}) \quad = \quad p(t^{lead}) \times p(t^{lag}) \tag{8}$$

3. *GHR Model:* In the GHR model, acceleration is used to decide whether to change a lane or not. This acceleration is computed using GHR [23] equation as follows,

$$a_n(t) \quad = \quad c v_n{}^m(t) \frac{\Delta v(t-T)}{\Delta x^l(t-T)} \tag{9}$$

where, $a_n$ is the acceleration of vehicle $n$ implemented at time $t$ by a driver and is proportional to, $v_n$ the speed of the $n^{\text{th}}$ vehicle, $\Delta v$ and $\Delta x$ are the speed and space spacing between the leader and subject vehicle. $c$ is the sensitivity co-efficient; $m$ is the speed exponent (-2 to +2), $l$ is the distance headway exponent (+4 to -1). We use $c = 15$, $m = 1$, and $l = 2$ for our simulation. Now, in the first step, if $a_n < 0$, the subject vehicle wishes to change the lane. Then, braking is computed using **Equation 9**, however, lane changing and gap acceptance decisions are taken similarly as Gipps' lane changing model.

## C. Vehicle Generation Model

The vehicles are generated according to the negative exponential distributions of vehicular headways. The probability density function is given by **Equation 10**.

$$f(x) \quad = \quad \lambda e^{-\lambda x} \tag{10}$$

From the above equation, the expression for exponential variate headway $X$ can be derived as:

$$X \quad = \quad \mu(-lnR) \tag{11}$$

where, μ is the mean headway, $R$ is the random number between 0 and 1.

### D. Both Lane and Non-Lane-Based Traffic Simulation

In our work, a link is composed of a number of strips. The width of the strips is a parameter of the simulator. The number of strips on a link is calculated as follows,

$$\# \text{ of strips} = \lfloor \frac{link\ width}{strip\ width} \rfloor$$

Each vehicle occupies some strips according to its width, computed as follows,

$$\# \text{ of occupied strips} = \lceil \frac{vehicle\ width}{strip\ width} \rceil$$

Now, if the strip width is greater than the width of a vehicle, the vehicle will be fully contained in a single strip and that single strip can be considered as a lane. Thus, by varying the strip width, we simulate both lane and non-lane-based traffic.

### E. Performance Metrics

To compute the overall performance of a road network under different parameters, we use four performance metrics as follows,

1. Average speed on a link ($\overline{speed_l}$ *km/hour*)
2. Average waiting time on a link ($\overline{t_l}$ *s*)
3. *Average vehicle flow rate of a link (vehicle/hour)*
4. Average speed of a vehicle ($\overline{speed_v}$ *km/hour*)

Now, we describe each of the performance metrics.

1. *Average speed on a link ($\overline{speed_l}$ km/hour)*: This metric represents the speed with which a vehicle typically crosses the corresponding link and is calculated for all links using **Equation 13**. Higher speed indicates better performance.

$$\overline{speed_{v_l}} = \frac{length_l}{time\_to\_cross_l} \tag{12}$$

$$\overline{speed_l} = \frac{\sum_{v \, \epsilon \, S_l} \overline{speed_{v_l}}}{|S_l|} \tag{13}$$

where, $\overline{speed_{v_l}}$ is the average speed of vehicle $v$ on link $l$ and $S_l$ is the set of all vehicles that crosses link $l$.

2. *Average waiting time on a link ($\overline{t_l}$ s)*: This metric indicates the average time a vehicle has to wait without movement in a link. It is calculated for all links using **Equation 14**. Lower average waiting time indicates better performance.

$$\overline{t_l} = \frac{\sum_{v=1}^{N} waiting\_time_v}{N} \tag{14}$$

where, $waiting\_time_v$ is the individual waiting time of vehicle $v$ while crossing link $l$ and $N$ is the total number of vehicles that leave link $l$.

3. *Average vehicle flow rate of a link (vehicle/hour):* This metric indicates the flow rate of a link and is calculated as the average number of vehicles that cross the middle of the corresponding link in an hour. The higher the better for this metric.

4. *Average speed of a vehicle ($\overline{speed_v}$ km/hour):* This metric represents the speed with which a vehicle travels and is calculated for all vehicles using **Equation 16**. Higher average vehicle speed indicates better performance.

$$\overline{speed_i} = \frac{total\_distance\_traveled_i}{total\_travel\_time_i} \tag{15}$$

$$\overline{speed_v} = \frac{\sum_{i \in S_v} \overline{speed_i}}{|S_v|} \tag{16}$$

where, $S_v$ is the set of all vehicles, $\overline{speed_i}$ is the average speed of the vehicle $i$, and $|S_v|$ is the total number of vehicles.

## EXPERIMENTAL SETUP

In this section, we describe the experimental setup of our comparative study.

### A. *Parameters of Our Simulation*

We vary a number of significant parameters of our simulator and gather data through simulations for further exploration and performance comparison among lane-based and non-lane-based road network.

1. *Topology:* We can give different road topology as input to our simulator through *node.txt*, *link.txt*, and *path.txt* file.

2. *Vehicle Generation Rate:* We have three different vehicle generation rates: low, medium, and high. To modulate the generation rate, we need to set the value of parameter *DemandType* as 0, 1, 2 for low, medium, and high generation rate respectively in *parameter.txt* file.

3. *Vehicle Distribution:* We have three types of vehicles: slow human-powered vehicles (max speed $\leq 15$ km/hour), vehicles with medium speed (30 km/hour $\leq$ max speed $\leq 50$ km/hour), and fast vehicles (80 km/hour $\leq$ max speed $\leq 120$ km/hour). We can set the percentage of the slow, medium, and fast vehicles using the parameters *SlowVehicle*, *MediumVehicle*, and *FastVehicle* of *parameter.txt* file.

4. *Strip Width:* This parameter is used to simulate lane-based or non-lane-based traffic behavior and can be set using *StripWidth* parameter of *parameter.txt* file or from the GUI of the simulator.

5. *Pedestrian Mode: Pedestrian Mode* is a single parameter. We can enable/disable the irregular crossing of roads by pedestrians by this parameter. To do this, we can change the value of *PedestrianMode* to *on/off* in the *parameter.txt* file or change it from the GUI.

### B. *Values of Our Parameter*

We conduct traffic simulation for 1800 s or half an hour for each sample and generate 10 samples using random number seed 1-10 for each scenario and take the average of the 10 samples for generating graphs and analyzing our result.

1. *Topology:* We use three road topologies extracted from the GIS map. They are parts of Dhaka, Miami, and Riyadh city shown in **Figure 2**.

2. *Vehicle Generation Rate:* We use 100, 400, and 800 vehicle/hour as low, medium, and high generation rate for Dhaka topology, respectively. On the other hand, we use 500, 1000, 2000 vehicle/hour as low, medium, and high generation rate respectively for Miami and Riyadh topologies.

3. *Vehicle distribution:* Properties and distribution of vehicles used in the simulations are shown in **Table 1**
4. *Strip width:* We use 0.5 m and 2.5 m as strip width to simulate non-lane and lane-based road traffic respectively.

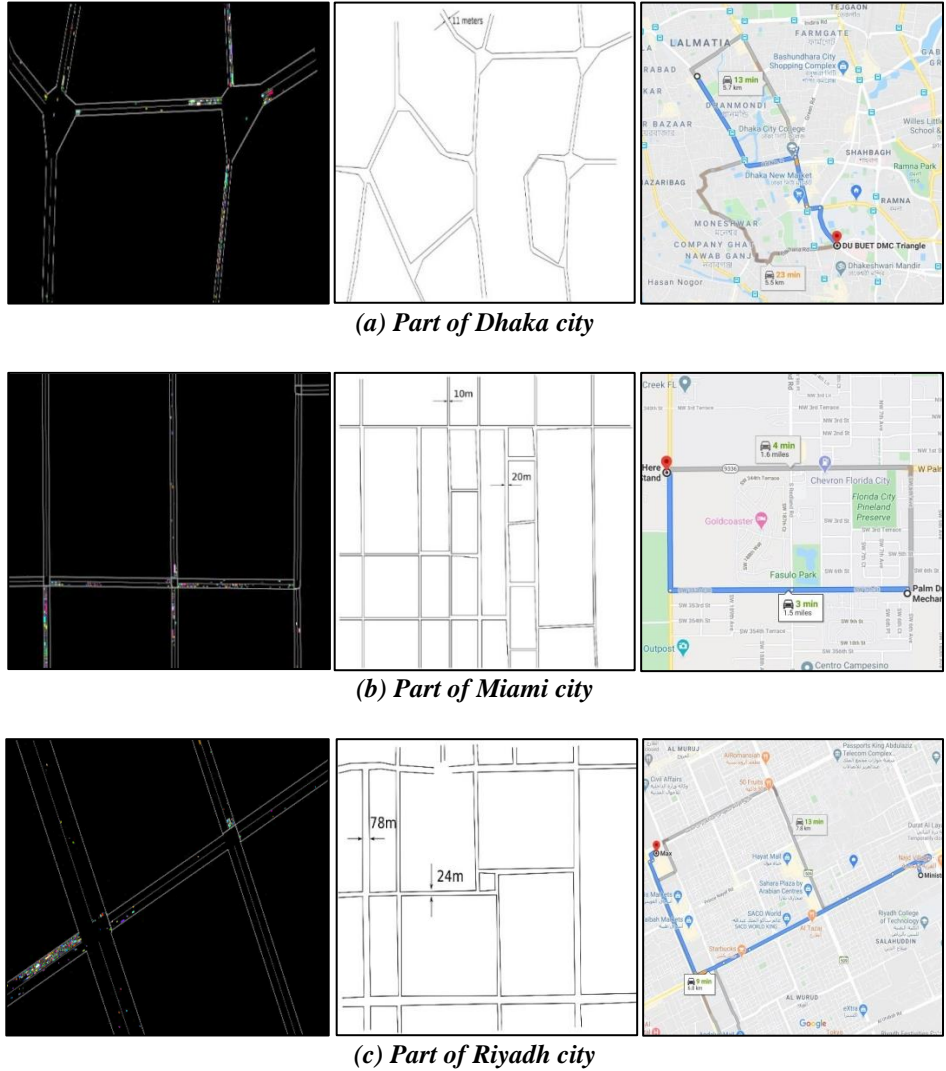

*(a) Part of Dhaka city*

*(b) Part of Miami city*

*(c) Part of Riyadh city*

*Figure 2: Simulation environment and topologies used in our simulations*

## VALIDATION OF ROADBIRD

Validation is an integral part of ensuring the credibility of a simulator, as it checks how accurately the simulator represents the real world. Accordingly, we perform validation of outcomes of RoadBird in comparison to real-world cases.

### A. Data Collection for Validation

In the process of our validation, we choose travel time as a measure of effectiveness (MOE). To measure the travel time, we select a route in the Dhaka city from Shankar to Palashi. To get real travel times on the selected route over several periods, we collect travel time data in both directions over the route for two types of vehicles namely cars and motorbikes. In the process of collecting travel time data, we log real-time travel times from Google Map, as it provides live travel time. Besides, we have extracted the selected route from the GIS map to run our simulations in RoadBird. We perform several iterations over the simulator to collect outcomes of several simulation runs.

***Table 1: Vehicle generation distribution***

| Speed Category | Vehicle Type | Vehicular Modal Share | Vehicle Distribution | |
|---|---|---|---|---|
| | | | Dhaka | Miami and Riyadh |
| Slow Vehicles | Bicycle | 9% | 55% | 9% |
| | Rickshaw | 89% | | |
| | Van/Cart | 2% | | |
| Medium Vehicles | CNG | 83% | 40% | 75% |
| | Bus (2 variations) | 15% | | |
| | Truck (2 variations) | 2% | | |
| Fast Vehicles | Motorbike | 88% | 5% | 16% |
| | Car (3 variations) | 12% | | |

***Table 2: Simulation parameters***

| Simulation time (minutes) | 40 | Vehicle generation rate (vehicle/hour) | Low traffic density | 100 |
|---|---|---|---|---|
| # of iterations in each case | 15 | | Medium traffic density | 400 |
| # of intersections | 8 | | High traffic density | 800 |
| # of vehicle generating nodes | 8 | Pedestrian Mode | | On |
| # of links | 18 | Strip Width (m) | | 0.5 |
| Distance between the two destinations (km) | 4.3 | | | |

## B. Experimental Setup for Validation

We use three different vehicle generation rates in our simulator. Besides, we divide the whole duration of a day (24 hours) into three parts based on on-road traffic density based on our day to day experience - 1) 10:00 PM - 09:00 AM for low traffic density, 2) 09:00 AM - 04:00 PM for medium traffic density, and 3) 04:00 PM - 10:00 PM for high traffic density. We have collected 63 real travel data from the Google Map in total over these durations in recent times. We do not use travel time data during the pandemic period when the lockdown period is in operation due to COVID-19, as traffic density is substantially less than usual during this period. Out of the 63-travel data from Google Map, we have collected 34 data during the pre-Corona pandemic age and the rest are during the Corona pandemic age. We have used some data from the Corona pandemic age as these data are applicable for low and medium traffic density. However, we collect high traffic density data only from the pre-Corona pandemic age.

Besides, in case of our simulations, we collect travel times from 45 iterations in total, where we consider 15 iterations for each of the three different traffic density cases. **Table 2** presents our simulation parameters.

## C. Performance Comparison

We present our simulation results in **Figure 3**. Here, **Figure 3(a)** and **3(b)** present average travel times of cars under low, medium, and high traffic density in both directions of the selected route. Besides, **Figure 3(c)** and **3(d)** present average travel times of motorbikes under low, medium, and high traffic density in both directions of the selected route. Results presented in **Figure 3** show that travel times obtained through simulations of RoadBird and obtained in real scenarios closely match with each other. To further dig into how far they match each other, we perform several statistical analyses over the travel time data. We present the outcomes of our statistical analysis next.

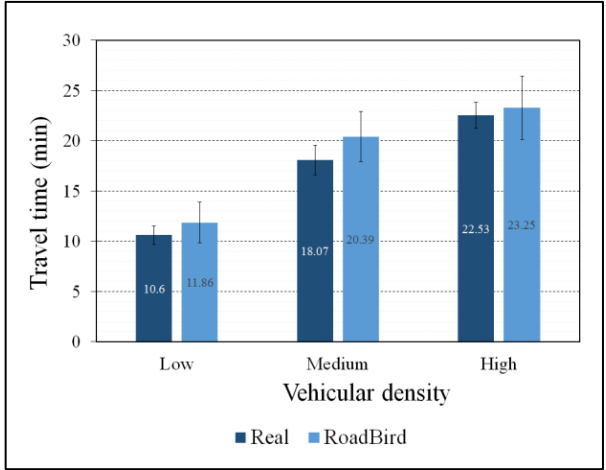

*(a) Travel time from Shankar to Palashi with car*

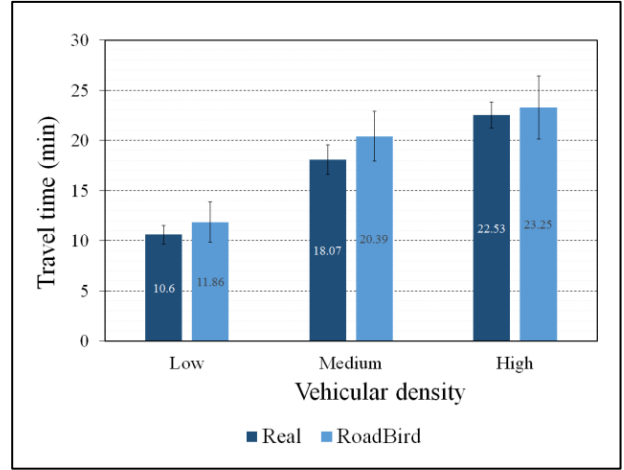

*(b) Travel time from Palashi to Shankar with car*

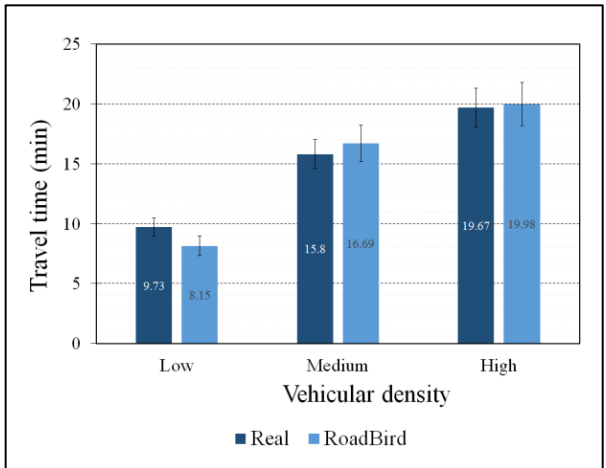

*(c) Travel time from Shankar to Palashi with motorbike*

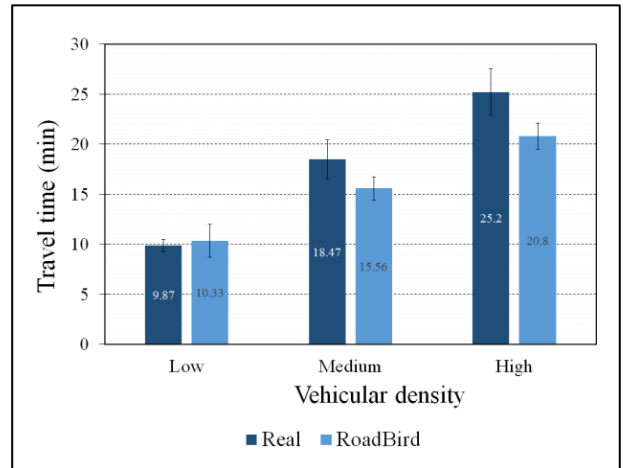

*(d) Travel time from Palashi to Shankar with motorbike*

**Figure 3: Travel time comparison between observed and simulated data for car and motor-bike in Shankar to Palashi route**

### D. Statistical Analysis over Real and Simulated Outcomes

For statistical validation, we consider a two-sample t-test and Kolmogorov-Smirnov (K-S) test with a 5% level of significance [24]. The null hypothesis in these tests is that the travel time observed from the real world and the travel time computed by our simulator come from the same distribution, and the alternative hypothesis is they come from different distributions. Then we compute the p-value. From the p-value, we can estimate how closely the simulation matches the real world. Goodness-of-fit measures are generally used to compute the overall performance of simulation models. According to Toledo et al., [25], Ni et al., [26], popular goodness-of-fit measures are, Mean Error (ME), Mean Absolute Error (MAE), Root Mean Square Error (RMSE), Mean Absolute Percentage Error (MAPE), and Root Mean Square Percentage Error (RMSPE), etc. We can quantify the overall error of our simulator with these statistics. So, we have computed these statistics and the results are summarized in **Table 3**.

**Table 3: Summary result for t-test and error computation**

| Route | Shankar to Palashi | | | | | | Palashi to Shankar | | | | | |
|---|---|---|---|---|---|---|---|---|---|---|---|---|
| Vehicle Type | Car | | | Motorbike | | | Car | | | Motorbike | | |
| Vehicle Density | low | medium | high | low | medium | high | low | medium | high | low | medium | High |
| p-value of t-test | 0.281 | 0.134 | 0.690 | *0.009* | 0.381 | 0.802 | 0.637 | 0.242 | 0.120 | 0.620 | *0.021* | *0.018* |
| p-value of K-S test | 0.375 | 0.181 | 0.375 | *0.009* | 0.626 | 0.925 | *0.003* | 0.181 | 0.181 | 0.375 | 0.076 | *0.009* |
| ME (min) | -1.261 | -2.325 | -0.712 | 1.581 | -0.894 | -0.318 | 0.447 | 1.861 | 2.842 | -0.459 | 2.907 | 4.403 |
| MAE (min) | 2.046 | 2.359 | 3.360 | 1.581 | 0.898 | 0.931 | 1.858 | 2.217 | 3.248 | 1.349 | 2.907 | 4.576 |
| RMSE (min) | 2.753 | 3.214 | 3.780 | 1.659 | 1.152 | 1.337 | 2.99 | 2.464 | 3.875 | 2.315 | 3.352 | 5.078 |
| MAPE (%) | 18.333 | 11.933 | 15.153 | 16.640 | 5.320 | 5.533 | 16.84 | 11.367 | 11.233 | 12.467 | 14.547 | 17.327 |
| RMSPE (%) | 24.199 | 15.314 | 17.380 | 17.611 | 6.476 | 8.692 | 25.72 | 12.811 | 12.994 | 19.780 | 16.056 | 18.353 |

### E. Qualitative outcomes of our statistical analysis

Here are the qualitative findings of our experiment:

1. From the p-value of both of our two-sample t-test and two-sample K-S test, with a 5% level of significance, we can see that in 9 out of 12 cases, we have p-value greater than our chosen LOS which clearly states that our simulation closely matches the real-world traffic scenario.
2. Mean Error (ME) indicates the existence of systematic under- or over-prediction in the simulated measurements. From our calculation, we can say that our simulator is not inherently biased in any direction.
3. From Mean Absolute Error (MAE) and Root Mean Square Error (RMSE), we can say that our predicted travel time varies from less than 1 minute to at most 5 minutes.
4. From Mean Absolute Percentage Error (MAPE), we can say that the accuracy of our simulator varies from 82-95% with an average accuracy of ~88%.
5. From Root Mean Square Percentage Error (RMSPE), we find that the average accuracy of our simulator is ~85% with a peak value of ~94%. Though RMSPE penalizes the outliers more, we have only two cases where the accuracy drops to ~75%.

## RESULTS

After validating our simulator, we simulate lane-based and non-lane-based traffic systems with the parameters described in the experimental setup part. Now, we present their comparative performance concerning different metrics.

### A. Performance Comparison for Dhaka Topology

As a representative city of the developing world, we consider Dhaka city to simulate both lane-based and non-lane-based traffic through varying different parameters of RoadBird. In the following subsections, we discuss the comparative performance of lane-based and non-lane-based traffic in Dhaka city.

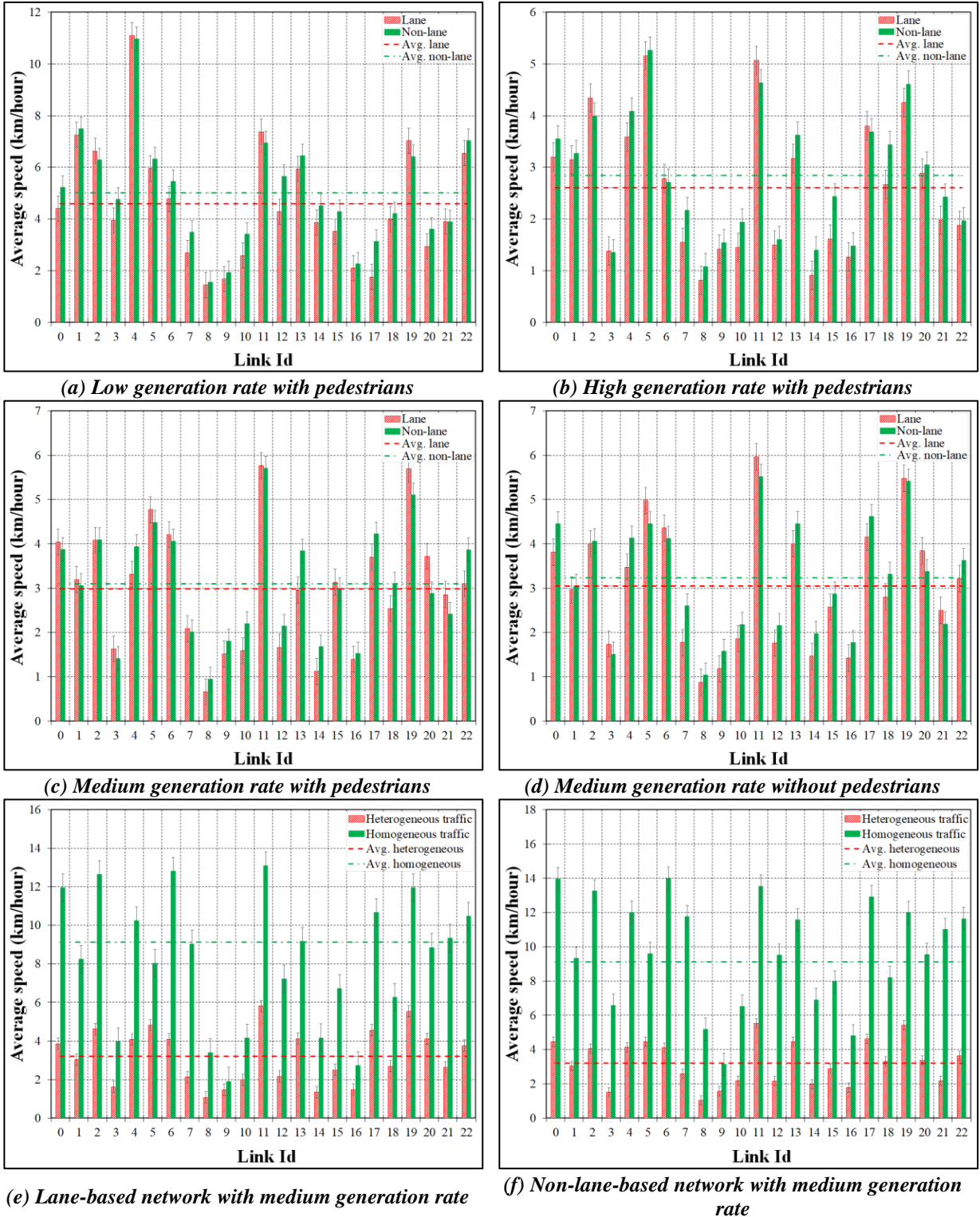

(a) *Low generation rate with pedestrians*

(b) *High generation rate with pedestrians*

(c) *Medium generation rate with pedestrians*

(d) *Medium generation rate without pedestrians*

(e) *Lane-based network with medium generation rate*

(f) *Non-lane-based network with medium generation rate*

*Figure 4: Performance comparison based on average speed on a link for Dhaka topology under various combinations*

1. *Performance comparison based on average speed on a link:* We present performance comparison based on average speed on a link for Dhaka topology under various combinations in **Figure 4**. According to this figure, the average speed on a link decreases with increasing vehicle generation rate. Besides, the average speed on links is below 10 km/hour that matches the World Bank report [1], i.e., the average speed of Dhaka city is around 7 km/hour. We can also see that irrespective of generation rate, the average speed for the non-lane road network is higher than that of lane-based road networks. As roads of Dhaka are narrow and most of the vehicles are small such as rickshaw, motorbike, car, etc., non-lane-based systems better utilize the road space. Besides, as the traffic is heterogeneous and slow vehicles prevail in high ratios, speed is mainly controlled by the slow vehicles. This is the reason for the higher speed in the non-lane-based traffic stream. Here, we enable the pedestrians and adopt heterogeneous vehicle distribution as shown in **Table 1**. If we compare the performance by keeping generation rate fixed to medium rate and enabling or disabling pedestrians on the road, the performance does not get changed significantly due to heterogeneous traffic. If we compare the performance by changing the distribution of vehicles into homogeneous, performance significantly improves for both lane-based and non-lane-based traffic as slow vehicles are removed from the traffic stream.

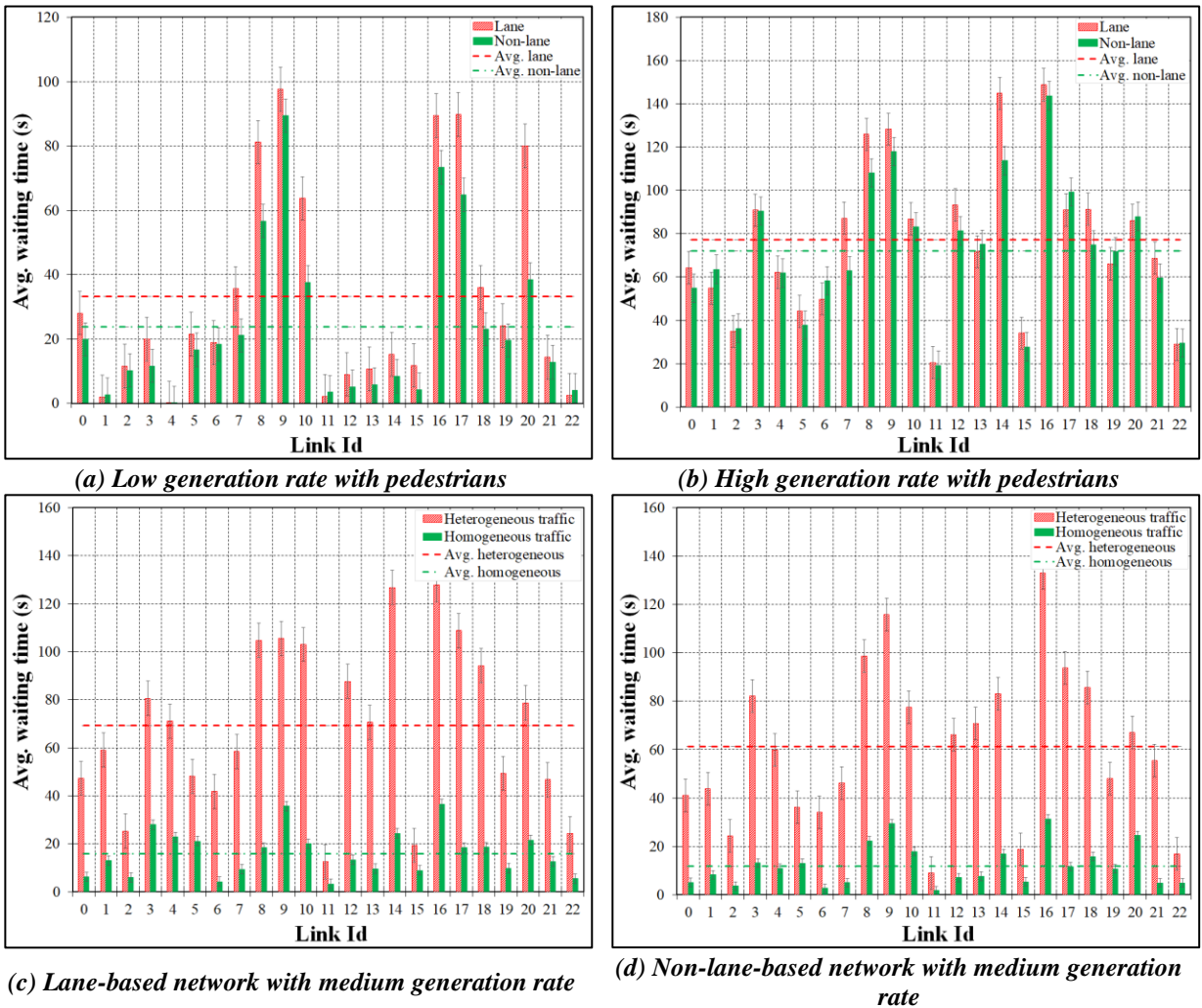

*(a) Low generation rate with pedestrians*

*(b) High generation rate with pedestrians*

*(c) Lane-based network with medium generation rate*

*(d) Non-lane-based network with medium generation rate*

**Figure 5: Performance comparison based on average waiting time on a link for Dhaka topology under various combinations**

2. *Performance comparison based on average waiting time on a link:* We depict performance comparison based on average waiting time on a link for Dhaka topology under various combinations in **Figure 5**. According to this figure, average waiting time increases with the increase in vehicle generation rate and average non-lane waiting time is less than lane-based waiting time. If we change the traffic distribution into homogeneous, waiting on links significantly reduces.

3. *Performance comparison based on average vehicle flow rate of a link:* We present a performance comparison based on the average vehicle flow rate of a link in **Figure 6(a)** and **6(b)**. From these figures, it is clear that the average vehicle flow rate increases with an increasing generation rate. Besides, the vehicle flow rate significantly increases for both lane-based and non-lane-based traffic networks when traffic distribution switches from heterogeneous to homogeneous as shown in **Figure 6(c)** and **6(d)**.

4. *Performance comparison based on the average speed of vehicle:* We present the comparative performance based on the average speed of vehicles in **Figure 6(e)** and **6(f)**. We can see that the average speed of vehicles decreases with the increase in vehicle generation rate. Besides, although non-lane-based network performs better in low generation rate, lane-based networks outperform non-lane-based network in high generation rate. This observation supports our practical experience. When the generation rate is high, the structured organization of lane-based traffic helps it to achieve better speed.

## B. Performance Comparison for Miami Topology

In this subsection, we present the comparative study between Dhaka and Miami traffic systems through simulating lane-based and non-lane-based traffic on Miami topology. Although non-lane-based traffic cannot be seen in Miami city, Miami's traffic system becomes lane-less during a hurricane evacuation. Hence, we try to know what will happen if Dhaka's unstructured traffic runs in Miami. Next, we present the comparison based on the following performance metrics for lane and non-lane-based traffic in Miami topology.

1. *Performance comparison based on average vehicle flow rate on the link:* We depict average vehicle flow rates on links in **Figure 7(a)** and **7(b)**. From these figures, it is clear that average vehicle flow rates for Dhaka distribution are higher than that of Miami distribution. The gap between the Dhaka traffic distribution and Miami traffic distribution becomes more prominent in case of the non-lane-based network (**Figure 7(b)**). Since most of the vehicles in Dhaka distribution are small and the roads of our chosen Miami topology are narrow as well, lane-less traffic maximizes road-space utilization, hence, it moves faster than structured traffic.

2. *Performance comparison based on average waiting time on the link:* Average waiting time on the link is shown in **Figure 7(c)** and **7(d)**. From these figures, it is clear that Dhaka's traffic system experiences less waiting time than Miami's one. This difference is more prominent in the lane-less network as before.

## C. Performance Comparison for Riyadh Topology

In this subsection, we compare performance between lane-based systems with homogeneous traffic and non-lane-based systems with heterogeneous traffic on Riyadh topology. **Figure 7(e)** and **7(f)** show the performance comparison between Dhaka's traffic (lane-less) and Riyadh's traffic (lane-based). Unlike Miami, lane-based traffic performs better in Riyadh in both vehicle flow rate and waiting time. Since roads of Riyadh are much wider and longer than Dhaka roads as shown in **Figure 2**, it can accommodate a lane-based system better.

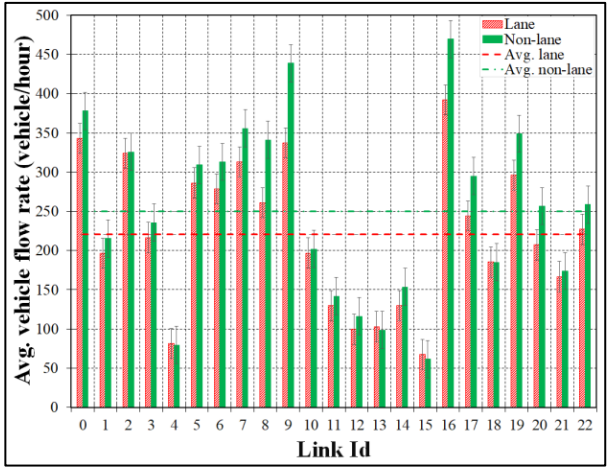

*(a) Average vehicle flow rate in low generation rate with pedestrians*

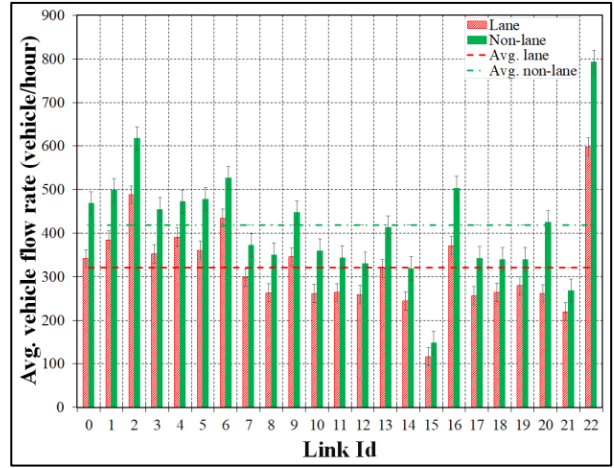

*(b) Average vehicle flow rate in high generation rate with pedestrians*

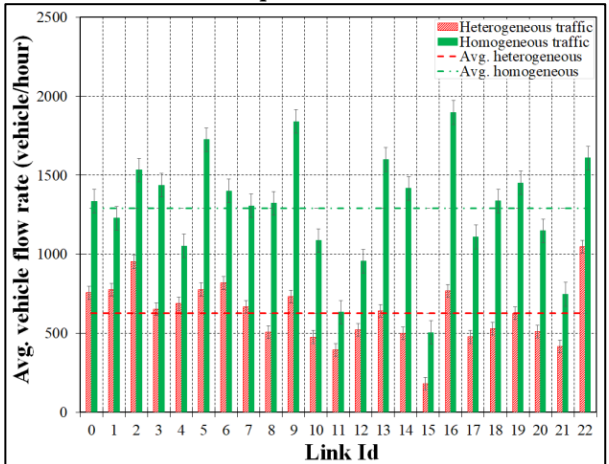

*(c) Average vehicle flow rate in lane-based network with medium generation rate*

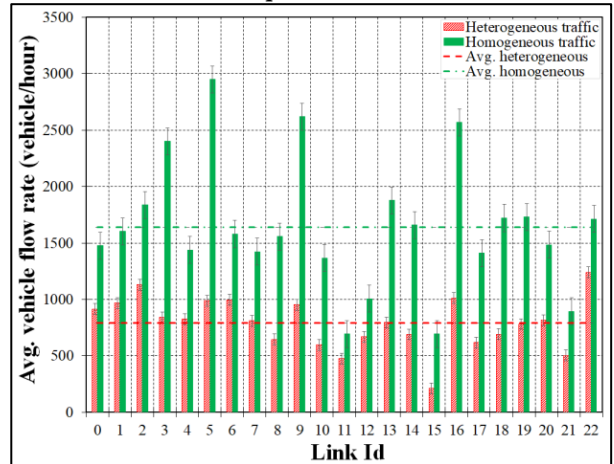

*(d) Average vehicle flow rate in non-lane-based network with medium generation rate*

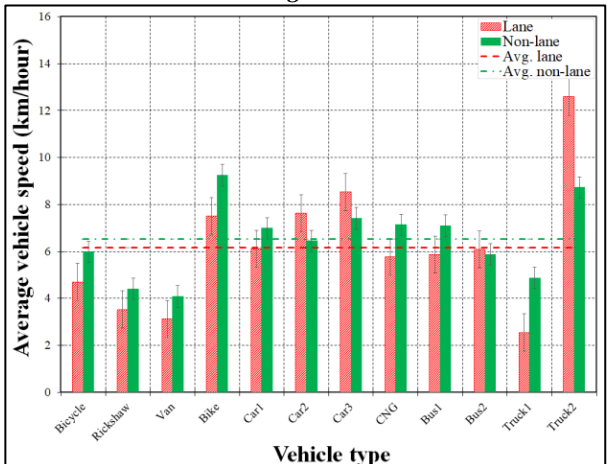

*(e) Average speed of the vehicles in low generation rate with pedestrians*

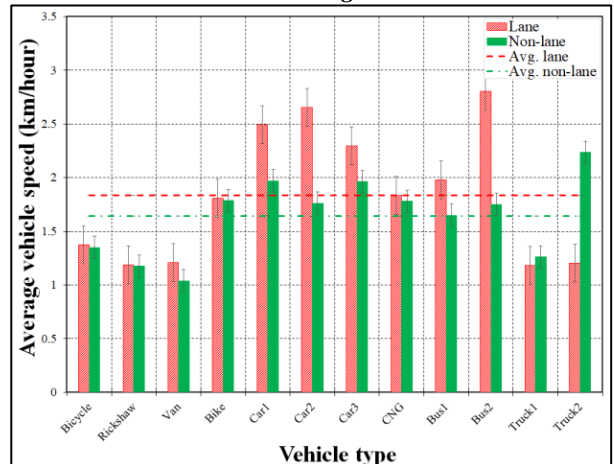

*(f) Average speed of the vehicles in high generation rate with pedestrians*

**Figure 6: Performance comparison based on the average vehicle flow rate of a link and average speed of the vehicle for Dhaka topology under various combinations**

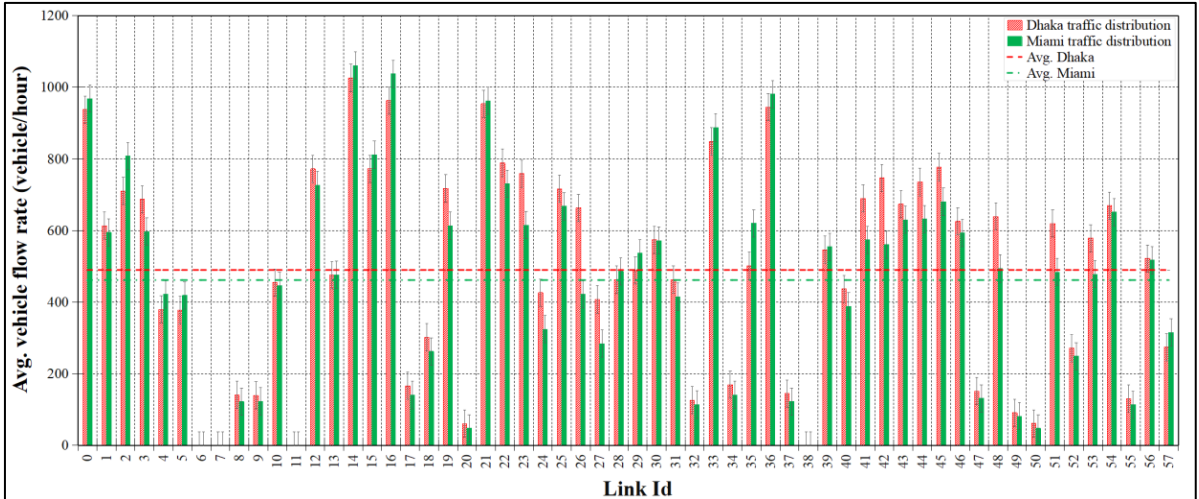

*(a) Average vehicle flow rate in the lane-based network for Dhaka and Miami traffic systems in Miami map*

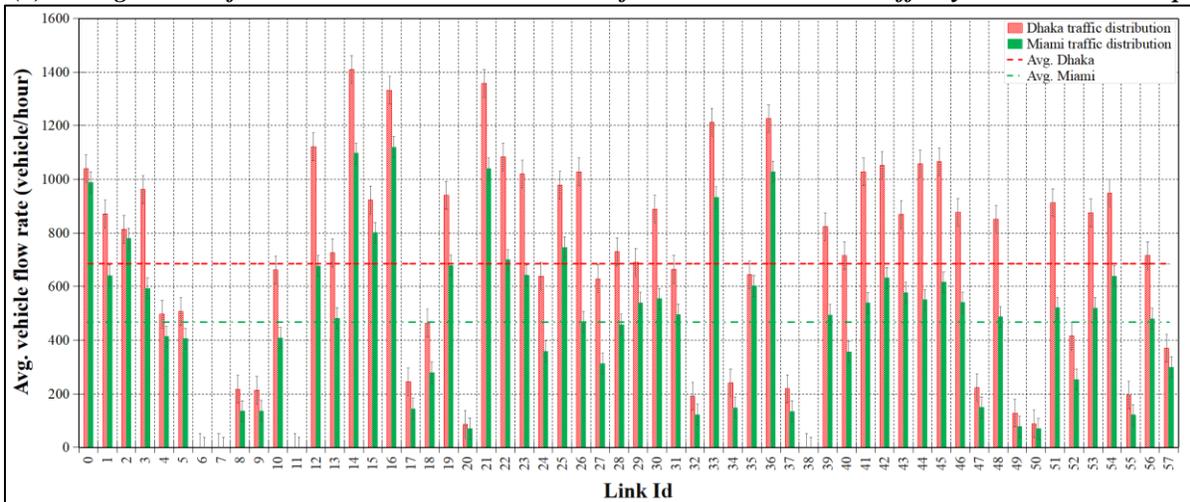

*(b) Average vehicle flow rate in the non-lane-based network for Dhaka and Miami traffic systems in Miami map*

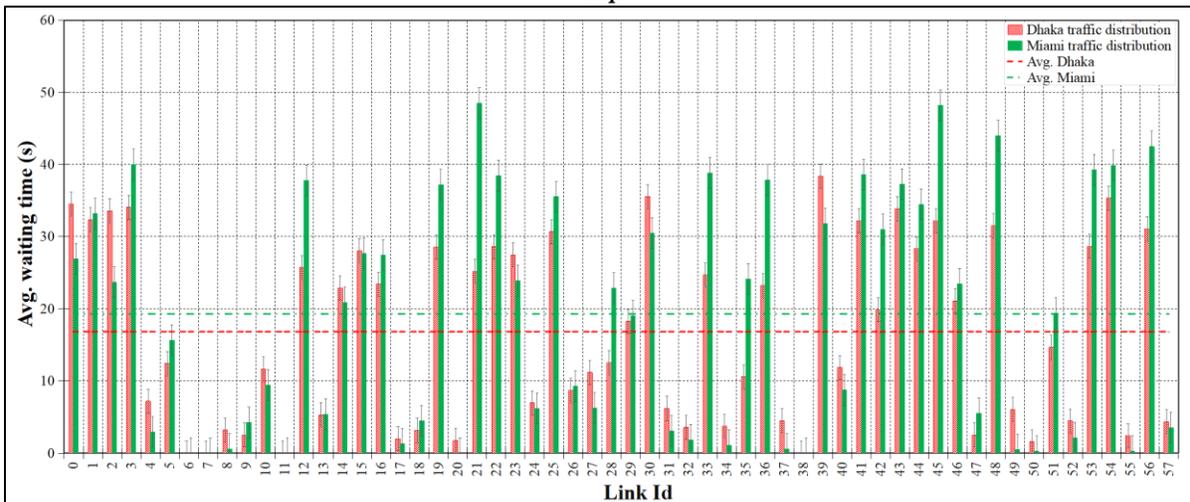

*(c) Average waiting time in the lane-based network for Dhaka and Miami traffic systems in Miami map*

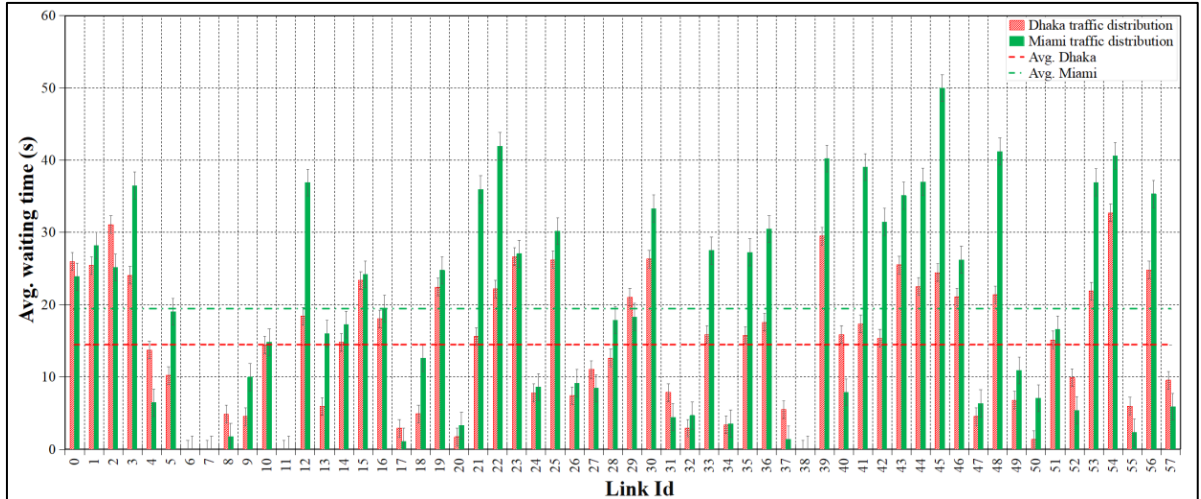

*(d) Average waiting time in the non-lane-based network for Dhaka and Miami traffic systems in Miami map*

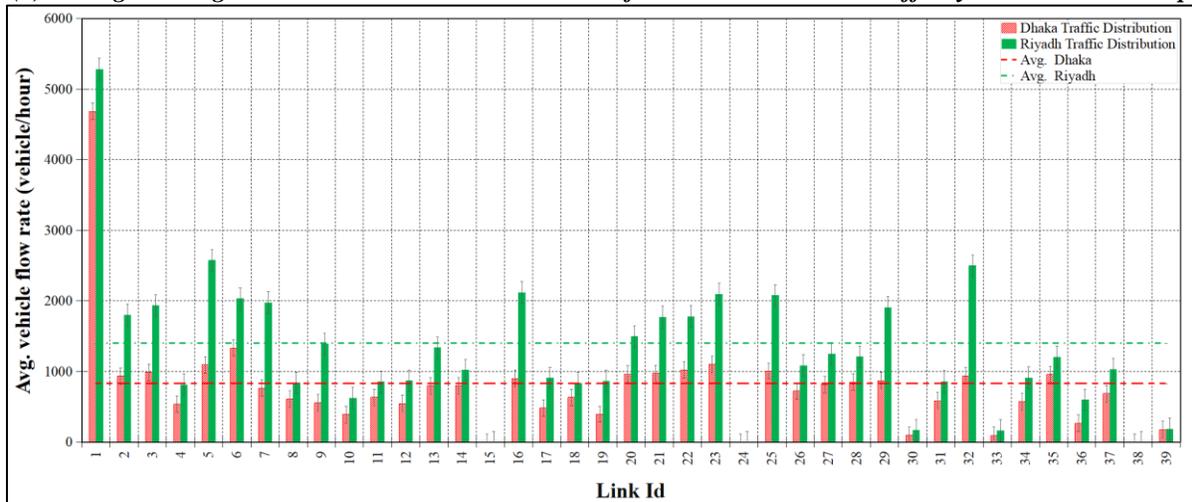

*(e) Average vehicle flow rate for Dhaka (non-lane) and Riyadh (lane) traffic systems in Riyadh map*

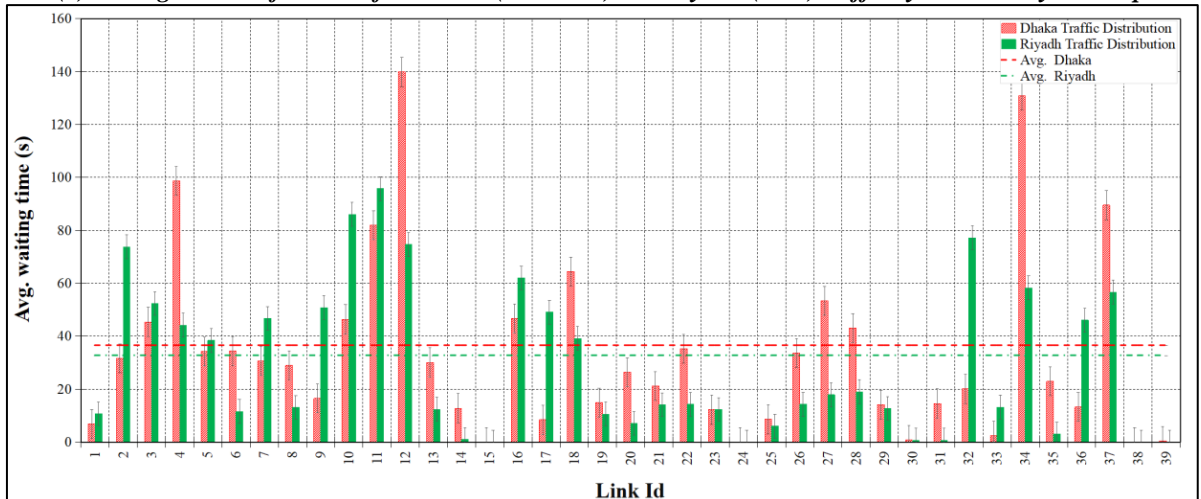

*(f) Average waiting time for Dhaka (non-lane) and Riyadh (lane) traffic systems in Riyadh map*

**Figure 7: Performance comparison among Dhaka, Miami, and Riyadh traffic systems for different performance metrics**

## CONCLUSION AND FUTURE WORK

The current traffic system in developing countries is not as well-disciplined as the ones in developed countries. Hence, a comprehensive comparative study of these systems has not received much attention from the academic community. There has been a major gap in the literature in the above-mentioned contexts and existing studies do not address most of the major issues in these scenarios. In our work, we study extensively the dilemma of setting up hypothetically lane-based traffic systems in an existing non-lane-based setup. The key findings of our comparative study can be described as follows:

- *The average speed on links and vehicle flow rate on links decreases with increasing vehicle generation rate for both lane and non-lane-based systems.*
- *Waiting time on links increases with increasing vehicle generation rates for both lane and non-lane-based systems.*
- *Irregular road crossing by pedestrians do not significantly alter the performance of heterogeneous traffic stream but slightly decreases the performance of homogeneous ones.*
- *With heterogeneous traffic, non-lane-based traffic systems perform better when roads are narrow.*
- *Lane-based traffic outperforms non-lane-based traffic when traffic stream is homogeneous and roads are wide enough to accommodate lane-based systems.*

From the above summary, we can say that only converting the traffic system of Dhaka into a lane-based system does not guarantee better performance due to the small, slow, human-powered vehicles, and narrow roads of Dhaka city. In addition, non-lane-based scenarios may provide better system performance in many cases such as emergency evacuations where one may expect a fair combination of slow and faster-moving vehicles. For example, lane reversal or contra-flow many times come into practice during major evacuations, however, such scenarios may benefit by adopting non-lane-based strategies. Moreover, RoadBird is capable of modeling behavioral phenomena such as pedestrians, bicycles, and signal control among others.

While we do not claim that establishing non-lane-based systems in all scenarios will eradicate the current problems overnight, we have clear empirical evidence that wide road networks can indeed benefit from lane-based systems. However, our experiments also show that narrower roads perform worse in lane-based systems. A natural future direction of our work would be to update our model with more complex and realistic parameters (e.g., traffic safety, road intersections, traffic signaling, pedestrians). Another plausible extension could be the usage of a combined setup of both lane and non-lane-based systems in the same topology. Moreover, such simulation techniques could potentially lead to new research directions of identifying more efficient strategies in major crises such as a hurricane or wildfire evacuation where demand typically exceeds capacity.

## ACKNOWLEDGEMENTS

The authors would like to acknowledge the Department of CSE, BUET for providing the HEQEP server to extract GIS maps and run simulations over the extracted maps. The authors also acknowledge FIU IRCC and its servers for running simulations.